\documentclass[twocolumn,superscriptaddress,prapplied]{revtex4-1}
\usepackage{graphicx}
\usepackage{amssymb}
\usepackage{amsmath}
\usepackage{epsfig}
\usepackage{color}
\usepackage{mathtools}
\usepackage[colorlinks,linkcolor=blue,anchorcolor=blue,citecolor=blue,urlcolor=blue]{hyperref}

\begin{document}
\title{High contrast quantum imaging with time-gated fluorescence detection}

\author{Xiang-Dong Chen}
\author{Yu Zheng}
\author{Bo Du}
\author{Deng-Feng Li}
\author{Shen Li}
\author{Yang Dong}
\author{Guang-Can Guo}
\author{Fang-Wen Sun}
\email{fwsun@ustc.edu.cn}

\affiliation{CAS Key Lab of Quantum Information, School of physics, University of Science and Technology of China, Hefei, 230026, P.R. China}
\affiliation{CAS Center For Excellence in Quantum Information and Quantum Physics, University of Science and Technology of China, Hefei, 230026, P.R. China}
\date{\today}
\begin{abstract}

Optical detection of spin state has been widely used for the solid state spin qubit in the application of quantum information processing. The signal contrast determines the accuracy of quantum state manipulation, sensitivity of quantum sensing and resolution of quantum imaging. Here, we demonstrated a time-gated fluorescence detection method for enhancing the spin state signal contrast of nitrogen vacancy (NV) center in diamond. By adjusting the delay between time gate and the excitation laser pulse, we improved both the signal contrast and signal-to-noise ratio for NV spin detection. An enhancement ratio of 1.86 times was reached for the signal contrast. Utilizing the time-gated fluorescence detection, we further demonstrated a high contrast quantum imaging of nanoparticle's stray magnetic field.
Without any additional manipulation of the quantum state, we expect that this method can be used to improve the performance of various applications with NV center.

\end{abstract}

\maketitle

\section{Introduction}
With stable optical properties and long spin coherence time, the negatively charged nitrogen vacancy (NV) in diamond has been widely studied in quantum information processing, biological labeling and nanophotonics. The detection of NV center spin state is among the most important techniques for applications of NV center in quantum computation\cite{jel-prl2004-1,wra-review-2006,Doherty2013phyrrevie}, quantum sensing\cite{lukin-nature2008-1,jel-nature2008-1,degen-review-RMP2017} and super-resolution imaging\cite{lukin-farfield,Walsworth-spinresolft-2017oe,Englund-nl2013-mw,gum-nanoscopy-lsa2017}. Though the electronic readout has been demonstrated\cite{Martin-ereadout-2017prl,Nesladek-ereadout-2015nc}  with NV center spin state, the optical fluorescence detection with high efficiency and simple setup is still the most widely used method. Usually, non-resonant excitation is utilized to pump the spontaneous emission of NV center\cite{ODMR-science1997,bassett-2018-spin,childress-readout-josab2016,budker-singlet-2010}, especially for the applications at room temperature. The spin-dependent fluorescence emission is resulted from the spin-selective intersystem-crossing (ISC) through metastable state. The probability of non-radiative ISC transition determines the limit of fluorescence contrast between different spin states. It subsequently affects the performance of NV based quantum sensing and optical super-resolution microscopy.

Several methods have been developed to improve the optical signal contrast of NV center spin state. The spin-to-charge readout method was based on the strong pumping of spin-dependent charge state conversion process\cite{lukin2014spincharge,bassett-2018-SCC,Meriles-spincharge-apl2018}. Multi-frequency synchronous manipulation was applied for pumping NV center with hyperfine structure\cite{Lich-apl-2018}. And the nuclear spins in diamond were used as ancilla spins for repetitive readout electron spin state of NV center\cite{lukin-readout-2009sci,wrachtrup-readout-2010prb}. In these methods, multiple laser and microwave pulses, or specialized external magnetic field were required. They affect the spin manipulation process of NV center, and make the application of NV more complicated.

In this work, we utilized the temporal-filtering technique for spin-dependent spontaneous emission detection, and subsequently improved the spin state signal contrast of NV center. A time gate that was synchronized to the excitation laser pulse was applied as the temporal filter for photon counting. Due to the difference between fluorescence lifetimes with different spin states, the signal contrast increased with the delay between time gate and excitation pulse. An enhancement ratio of approximate 1.86 times was obtained for the spin state detection, while the signal-to-noise ratio (SNR)  was also improved. Using this time-gated fluorescence detection technique, we improved the  contrast for the magnetic field quantum imaging of nano particles. The results indicated that our method was effective for a variety of applications with NV center spin state, as it did not need any additional manipulation of the quantum state.

\begin{figure}
  \centering
  \includegraphics[width=8.5cm]{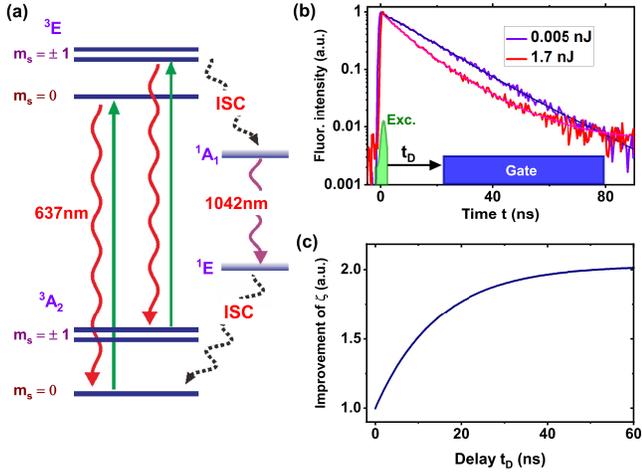}
  \caption{(a) The level scheme for NV center spin state detection. The laser excitations are indicated as solid arrows, while the solid curve arrows present the spontaneous emission.(b) The fluorescence decay traces of single NV center pumped by pulsed laser with different pulse energies. The bottom illustrates the time gate with delay of $t_{D}$. The lifetime of excited state is obtained through exponential functions fitting. (c) The calculated spin state signal contrast changed with $t_{D}$. }\label{figconcept}
\end{figure}

\section{Improve the signal contrast of spin detection}
The schematic of NV center energy level structure is shown in Fig. \ref{figconcept}(a). The ground state ($^{3}A_{2}$) and excited state ($^{3}E$) of NV are spin triplet states. A zero-field splitting, approximately 2.87 GHz at room temperature, is observed between  $m_{s}=0$ and $m_{s}=\pm 1$ in the ground state. There exists two intermediate spin singlet states ($^{1}A_{1}$ and $^{1}E$)\cite{budker-singlet-2010}. As mentioned before, the ISC through metastable singlet states shows significantly spin dependence. The ISC transition between $m_{s}=\pm 1$ in excited state and singlet state $^{1}A_{1}$ has a probability higher than that of the transition between $m_{s}=0$  in excited state and singlet state. And the $^{1}E$ metastable state preferentially decays to the $m_{s}=0$ in ground state. Consequently, the ISC transition initializes the spin state of NV to $m_{s}=0$. The non-radiative ISC transition also decreases the probability of spontaneous emission (zero photon line at 637 nm) from $m_{s}=\pm 1$ in the excited state. Though the infrared spontaneous emission (zero phonon line at 1042 nm) has also been observed with the transition between $^{1}A_{1}$ and $^{1}E$, it is much weaker than the visible emission\cite{budker-singlet-2010,manson-IR-2008njp}, and can be neglected. Therefore, the total fluorescence intensity with $m_{s}=\pm 1$ is lower than that with $m_{s}=0$ spin state. This is the base of optical detection of NV spin state.

Once the NV center is pumped to the excited state, the time-resolved photon emission probability of NV can be depicted as:
\begin{equation}\label{lifeitme}
  P_{i}(t)=P_{i}(0)e^{-t/\tau_{i}},
\end{equation}
where $\tau_{i}$ was the excited state's lifetime with spin state $m_{s}=i$.  As shown in Fig. \ref{figconcept}(a), the ISC transition from excited state to $^{1}A_{1}$ singlet serves as an additional decay channel for the $m_{s}=\pm 1$ spin state. As a result, the lifetime of excited state with  $m_{s}=\pm 1$ is shorter than that with $m_{s}=0$. In order to experimentally extract the information of fluorescence lifetimes with different spin states, we used pulsed laser with high power to pump the NV center. Due to the spin depolarization during charge state conversion process\cite{chen20152prb}, the time-resolved fluorescence emission with strong laser pumping is a bi-exponential decay, which is the combination of  emission from both $m_{s}=0$ and $m_{s}=\pm1$. In contrast, the single exponential decay is observed with weak laser pumping, mainly determined by the emission from $m_{s}=0$. The result in Fig. \ref{figconcept}(b) showed that the fluorescence lifetimes of NV were estimated to be $\tau_{0}\approx14 ns$ and $\tau_{\pm 1}\approx7.1 ns$, in consistent with previous report\cite{Awschalomprx600k}. It should be noted that, the exact values of lifetimes would be mediated by the environment and differ from NV centers..

From Fig. \ref{figconcept}(b), we can see that the difference of fluorescence emission between $m_{s}=0$ and $m_{s}=\pm 1$ changed with the decay time. For theoretical analyzing, we assume the photons in a time gate $t \in [t_{D},\infty)$ are counted for spin state estimation, where $t_{D}$ presents the delay between time gate and excitation laser pulse. Then, the detected fluorescence intensity can be written as:
\begin{equation}\label{intefluo}
  I_{i}(t_{D})=\int_{t_{D}}^{\infty}P_{i}(t)dt=P_{i}(0)\tau_{i}e^{-t_{D}/\tau_{i}}.
\end{equation}
Because the non-resonant excitation process is spin-independent, the initial emission probability is treated as $P_{0}(0) = P_{\pm 1}(0)$ here.  As the electron spin state is efficiently polarized to $m_{s}=0$ with laser pumping, the spin state signal contrast can be simply presented as:
\begin{equation}\label{eqcontrast}
  \zeta=\frac{I_{0}-I_{\pm 1}}{I_{0}}=1-\frac{\tau_{\pm 1}}{\tau_{0}}\cdot e^{-t_{D}(\frac{1}{\tau_{\pm 1}}-\frac{1}{\tau_{0}})}.
\end{equation}
It indicates that the spin state detection contrast increases with the delay $t_{D}$. The maximum signal contrast is $\lim \limits_{t_{D}\to \infty} \zeta (t_{D})=1$.

In previous works for NV optical detection, no time gate was applied for measuring the spontaneous emission. All fluorescence photons arrived in the interval of $t \in [0,\infty)$ were detected. Then the contrast for spin state detection without time gate would be $\zeta=1-\frac{\tau_{\pm1}}{\tau_{0}}$.  For the spin-dependent fluorescence lifetimes estimated in Fig. \ref{figconcept}(b), it means that the contrast $\zeta$ can be improved approximately two times by applying time gate for spin detection, as shown in Fig. \ref{figconcept}(c).

To experimentally demonstrate the enhancement of spin state signal contrast, we used a home-built confocal system for the NV center spin manipulation and detection. NV center in bulk diamond was produced through ion implanting. A 532 nm picosecond pulsed laser with a repetition rate of 10 MHz and average power of 0.14 mW was used to pump the fluorescence of NV center. A single photon counting modulator (SPCM) was used to detect the fluorescence of NV after passing through a long pass optical filter (edge wavelength 668 nm). The TTL pulses from SPCM were then selected by a FPGA card, and subsequently counted by a data acquisition card. The time gate and the excitation laser pulse were synchronized by an arbitrary function generator. The whole electronic setup was designed to count the TTL pulse with rising edge in a time region of $[t_{D},t_{D}+\Delta t)$.

\begin{figure}
  \centering
  \includegraphics[width=8.5cm]{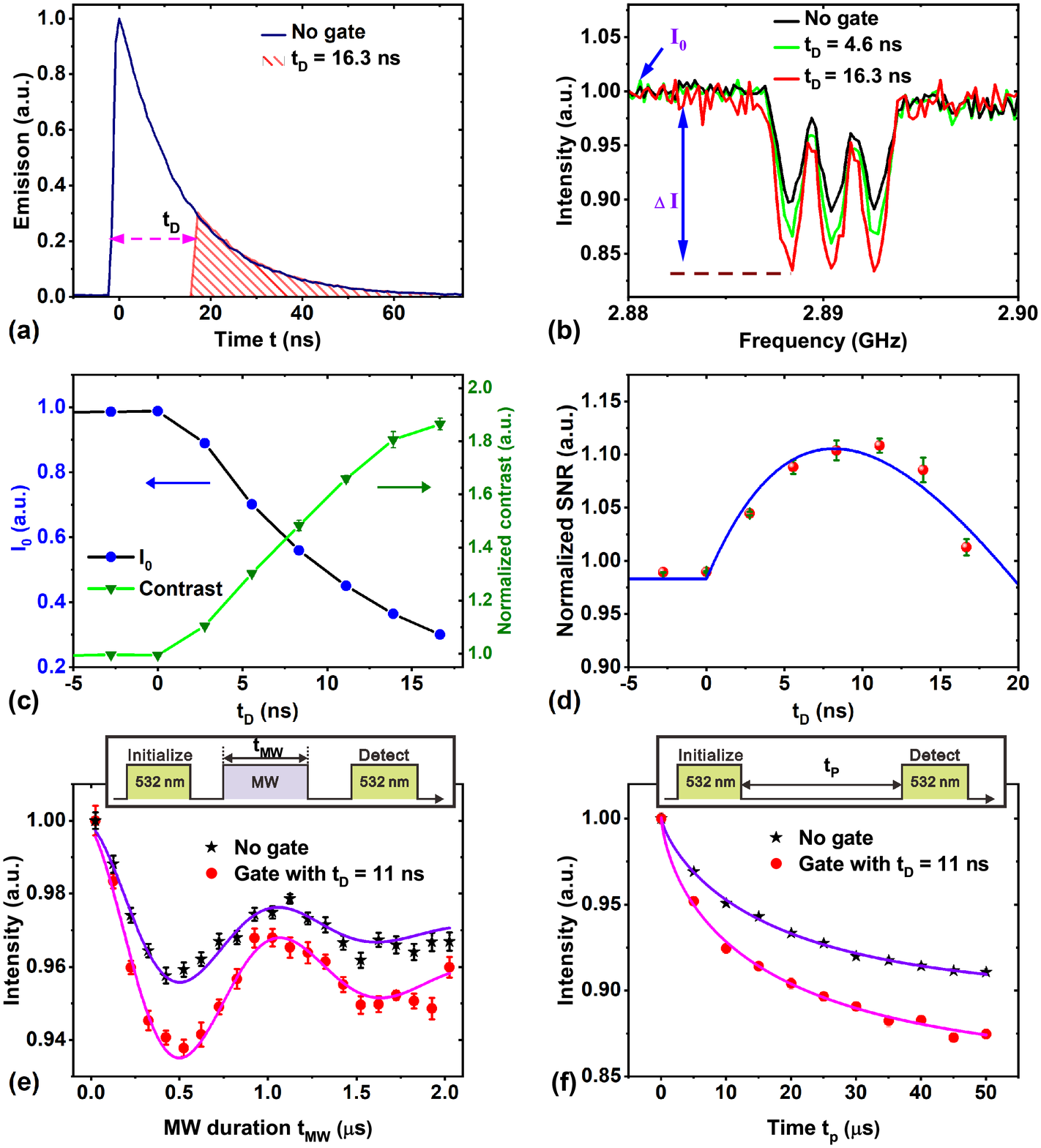}
  \caption{(a) The time-gated florescence decay traces. (b) The ODMR results of single NV detected with different temporal filtering.(c) The spin state signal contrast and the total fluorescence intensity changed with the delay of time gate. (d) The delay dependent SNR, which was calculated by using the values of contrast and $I_{0}$ in (c). Solid line was the fitting with Eq. \ref{eqSNR2}. (e) Rabi oscillation of NV center. (f) $T_{1}$ spin relaxation of NV center. The width of time gate was fixed at 55 ns here.}\label{figcontrast}
\end{figure}

To confirm the temporal-filtering performance of FPGA card, the time-resolved photon emission of single NV center was measured in Fig. \ref{figcontrast}(a). A time-correlated single photon counting module was used to record the TTL pulses those were selected by FPGA card. Here, a width of $\Delta t$ = 55 ns was applied for the time gate. We can see that, the photon emission in intervals of $[0,t_{D})$ and $[t_{D}+\Delta t,\infty)$ was perfectly blocked, while the emission during $[t_{D},t_{D}+\Delta t)$ was detected without any intensity reduction.

The optically detected magnetic resonance (ODMR) of NV electron spin was measured to quantify the change of signal contrast. After the spin state of NV was initialized to $m_{s}=0$, a microwave pulse was applied. We scanned the frequency of microwave. When it was resonant with the transition between $m_{s}=0$ and $m_{s}=\pm 1$ in ground state, the microwave will pump the NV center to $m_{s}=\pm 1$. Then there would show a decrease of the fluorescence intensity. The spin state signal contrast can be presented by $\zeta=\Delta I/I_{0}$, where $\Delta I$ is the amplitude of ODMR dip, as shown in Fig. \ref{figcontrast}(b). The experimentally measured value of contrast was lower than that deduced from Eq. (\ref{eqcontrast}). It might be resulted from the unperfect spin manipulation, spin polarization during readout process and the hyperfine structure of NV center.

Comparing the ODMR results with different time gates, we saw that the signal contrast can be significantly improved. Normalized by the value without time gate, the signal contrast was plotted as the function of delay $t_{D}$ in Fig. \ref{figcontrast}(c). As expected, the contrast of spin detection increased with the delay of time gate, while the width of gate was fixed at 55 ns. An enhancement ratio of approximate 1.86 was obtained with delay of 16 ns.

However, it should be noted that the total fluorescence intensity was decreased by increasing the delay of time gate, as shown in Fig. \ref{figcontrast}(c). This would affect the SNR of spin detection, which is important for enhancing sensitivity of quantum sensing. The SNR is presented as:
\begin{equation}\label{eqSNR}
  SNR=\frac{\Delta I}{I_{noise}}=\zeta \times \frac{I_{0}}{I_{noise}}.
\end{equation}
The noise depends on the square root of total photon counts. Then, the noise ratio is considered as $\frac{I_{noise}}{I_{0}}\propto \frac{1}{\sqrt{I_{0}}}$. The SNR can be written as:
\begin{equation}\label{eqSNR2}
 SNR \propto \zeta \sqrt{I_{0}} = \frac{\tau_{0}e^{-t_{D}/\tau_{0}}-\tau_{\pm 1}e^{-t_{D}/\tau_{\pm 1}}}{\sqrt{\tau_{0}}e^{-t_{D}/2\tau_{0}}}
\end{equation}
Using the experimental results in Fig. \ref{figcontrast}(c), the normalized SNR of spin detection was shown as the function of $t_{D}$ in Fig. \ref{figcontrast}(d). For the sample presented here, we found that the highest SNR was obtained with a delay of around 11 ns. The maximum SNR was enhanced by a factor of 1.1 comparing to that without time gate. Considering both contrast and SNR, we chose a delay of 11 ns for the detection in the rest of this work.

As mentioned before, our method did not rely on the manipulation of NV center spin state or charge state. It can be applied to all the applications with NV center spin state. In Fig. \ref{figcontrast}(e)(f), we presented the results of Rabi oscillation and spin relaxation. The whole pulse sequence of Rabi oscillation was several microsecond in length, while the length of pulse sequence for spin relaxation was up to 50 $\mu s$. By applying a time gate for spin detection, both the Rabi oscillation and spin relaxation signal contrasts were improved, while the dynamics of spin state was unchanged. It demonstrated that our method could be used at various situations. In contrast, the spin-to-charge readout technique is more effective for the pulse sequences with long lengths, as hundreds of microseconds or several milliseconds\cite{lukin2014spincharge,bassett-2018-SCC}.

\section{High contrast magnetic field imaging}

\begin{figure*}
  \centering
  \includegraphics[width=15cm]{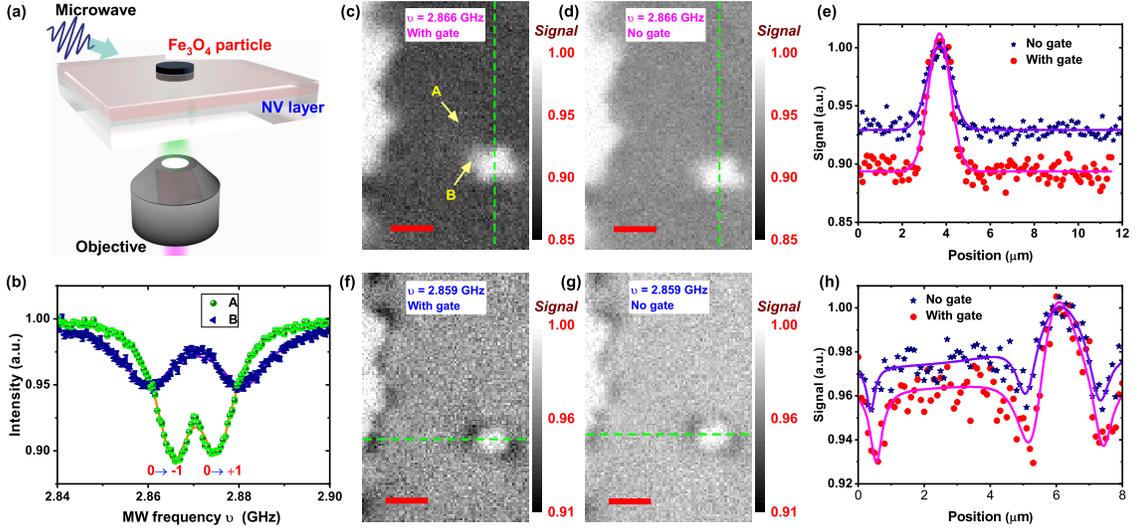}
  \caption{(a) The experimental schemtic for magnetic field imaging with NV center ensemble. (b) The ODMR results of NV ensembles at different positions. Two peaks corresponding to $m_{s}=0\rightarrow\pm1$ transitions were observed. As indicated in the magnetic field contour images, the spin states of NV centers at position $\textbf{B}$ were affected by the stray magnetic field of Fe$_{3}$O$_{4}$, while the NV centers at position $\textbf{A}$ only affected by the background magnetic field. (c)(d)The  contour images of magnetic field corresponding to $B_{z}=1.4 G$ ($\nu = 2.866$ GHz). (e) The cross-sections  of the dashed lines in (c)(d). (f)(g)The  contour images of magnetic field corresponding to $B_{z}=3.9 G$ ($\nu = 2.859$ GHz). (h) The cross-sections  of the dashed lines in (f)(g). Scale bars in the contours images: 2 $\mu m$. Pixel dwell time: 0.1 s.}\label{figODMRimaging}
\end{figure*}

The signal contrast is a key parameter for the microscopy imaging\cite{holl-2012-screp,holle2015magimaging,Tetiennee-sciadv2017-current}. And it can improve the spatial resolution of super-resolution microscopy\cite{hell-gate-2011NC}. For quantum sensing, the contrast is important for extracting the information of samples from the noisy background. In order to show how the temporal-filtering affected the results of quantum imaging, we used NV center to detect the magnetic field of  nano particles.

Here, the sample was synthesized Fe$_{3}$O$_{4}$ nano disk with diameter of approximate 200 nm. The particles were deposited on the surface of diamond plate, in which high density NV center ensembles have been produced through ion implanting, as shown in Fig. \ref{figODMRimaging}(a). The density of NV center array was estimated to be approximate $48/(100 nm)^{2}$. And the depth of NV center, which was determined by the energy of ion implanting, was 20 nm in average. The NV center was optically excited and detected from the backside of diamond plate.

Due to strong magnetic interaction, these Fe$_{3}$O$_{4}$ particles spontaneously formed the aggregates with size of several micrometers. The magnetic field of random placed Fe$_{3}$O$_{4}$ particles had two effects on the spin dynamics of NV center. The static component of magnetic field shifts the resonant frequencies of NV spin state transitions\cite{jayich-2016spm,holle2015magimaging,jacques-sensing-nc2013,yacoby-imaging-nn2012}, and fluctuating component reduces the coherence time of spin state\cite{hollenberg-nl-15,park-t1sensing-2014nl,wra-t1sensing-2015nl,jacques-relaxometry-2013prb,jayich-relaxometry-2014prappl}. Firstly, we measured the static component of magnetic field. As shown in Fig. \ref{figODMRimaging}(b), the ODMR results of NV center were changed by the local magnetic field of Fe$_{3}$O$_{4}$ particles. Due to Zeeman effect, the shifts of resonant frequencies can be simply written as $\Delta \nu = \pm \gamma_{e} B_{z}$ for transitions between $m_{s}=0$ and $m_{s}=\pm 1$, respectively. Here $\gamma_{e}=2.8 MHz/G$ is the electron gyromagnetic ratio and $B_{z}$ is the projection of magnetic field along NV axis.

In Fig. \ref{figODMRimaging}, the contour images of stray magnetic field were obtained by scanning the samples while pumping the NV center spin state transition with a frequency-fixed microwave pulse. The detected fluorescence intensity was then normalized by the result without microwave pulse, as $Signal=F_{MW}/F_{no MW}$. The contours images with two microwave frequencies $\nu=$ 2.866 GHz and 2.859 GHz were presented in Fig. \ref{figODMRimaging} (c)(d) and (f)(g), corresponding to the stray magnetic field with $B_{z}=$ 1.4 G and 3.9 G, respectively. The dark features in these images indicated the contours of magnetic field.

\begin{figure*}
  \centering
  \includegraphics[width=15cm]{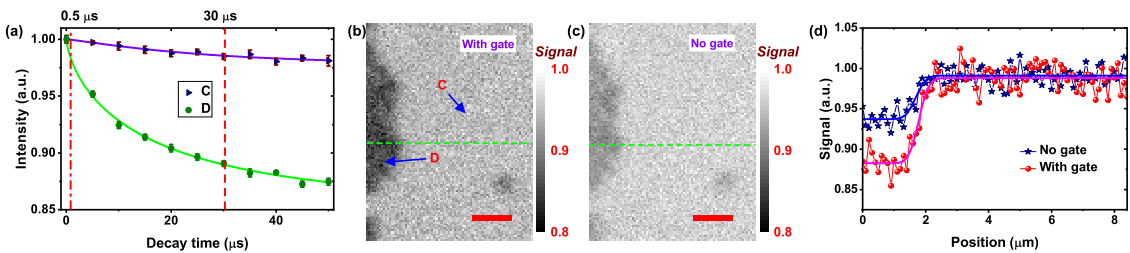}
  \caption{(a) $T_{1}$ spectroscopy of NV center with and without the affecting of the Fe$_{3}$O$_{4}$ magnetic field. The positions of relaxation measurements were indicated as C and D in the relaxometry image. (b)(c) The relaxometry images of NV center with and without time gate for spin detection. (d) The cross-sections corresponding to the dashed lines in (b)(c). Scale bars in the relaxometry images: 2 $\mu m$. Pixel dwell time: 0.2 s.}\label{figT1imaging}
\end{figure*}

A time gate with $t_{D} = 11$ ns and $\Delta t =$ 60 ns was applied for the magnetic field contour imaging, while no additional changes of the laser or microwave pulses was needed. Comparing the results with and without temporal filtering, we can see that the contrast was improved with the time gate. The contour lines can be better distinguished from the background. To further show the change of signal contrast, the cross-sections were presented in Fig. \ref{figODMRimaging}(e)(h). It showed that the contrast of magnetic field contour lines improved approximately 1.65 times with a time gate of $t_{D}=11 ns$, in consistent with the results in Fig. \ref{figcontrast}. The results with different microwave frequencies indicated that the time gate was effective for different contours imaging.

To gain the information of the stray magnetic field fluctuations of Fe$_{3}$O$_{4}$ aggregates, the two dimensional relaxometry of NV center was measured. The laser pulses same as that in Fig. \ref{figcontrast}(f) were applied to detect the $T_{1}$ spin relaxation of NV centers at different positions. The change of $T_{1}$ induced by magnetic field fluctuation was confirmed in Fig. \ref{figT1imaging}(a). The experimental decay traces were fitted by a single exponential decay function $F(t_{P})= F(\infty)+\Delta F \times  e^{-t_{P}/T_{1}}$. The results indicated that the presence of Fe$_{3}$O$_{4}$ aggregates changed the longitudinal coherence time $T_{1}$ from approximate $280 \mu s$ to 12.8$\mu s$.

We recorded the fluorescence intensities $F(t_{P})$ at two different decay times $t_{P}=30 \mu s$ and $t_{P}=$ 500 ns. The spin relaxometry images were then obtained by plotting the normalized fluorescence intensity $Signal=F(30 \mu s)/F(500 ns)$. A time gate ($t_{D} = 11$ ns, $\Delta t =$ 60 ns) was used for the relaxometry imaging. The same Fe$_{3}$O$_{4}$ aggregates in the magnetic contour images were detected here. As presented in Fig. \ref{figT1imaging}(b)(c), the features of spin relaxometry images agreed with the distribution of Fe$_{3}$O$_{4}$ aggregates in Fig. \ref{figODMRimaging}.  In consistent with the results of magnetic field contour imaging, the contrast of $T_{1}$ imaging was also improved by utilizing the time-gated fluorescence detection, as depicted in the cross-sections in Fig. \ref{figT1imaging}(d).

The pulse for contour images was short, with a length of approximate 1 $\mu s$ in Fig. \ref{figODMRimaging}. In contrast, the pulse for spin relaxometry imaging was longer, with a length of 30 $\mu s$. The results of magnetic field contour images and spin relaxometry images further proved that, the time-gated fluorescence detection can be effectively applied for a wide range of applications with NV spin.

\section{Discussion and conclusion}
Our results provided a simple and efficient way to improve the contrast for spin detection. To implement the time-gated fluorescence detection, we only need to update the data acquisition devices of a confocal microscopy system. For the super-resolution microscopy based on spin manipulation, such as spin-RESOLFT microscopy\cite{lukin-farfield,Walsworth-spinresolft-2017oe}, we expect that our method can further improve the spatial resolution of microscopy\cite{hell-gate-2011NC}. In addition, the time-gated detection would also enhance the contrast of NV center anomalous saturation effect induced by spin depolarization\cite{chen20152prb}, and subsequently can be used to improve the performance of ground state depletion nanoscopy\cite{hell-gsd-2012njp}. Our method can also be used for the non-resonant optical detection of other quantum emitters. As shown in Eq. \ref{eqcontrast}, the improvement of contrast would be more effective for the emitters with smaller difference between fluorescence lifetimes with different spin states. For instance, to detect the spin state of NV at high temperature\cite{Awschalomprx600k}, the time-gated fluorescence detection would be necessary.

In summary, we proposed and demonstrated the time-gated fluorescence detection for spin state estimation of NV centers. It showed that the fluorescence contrast between $m_{s}=0$ and $m_{s}=\pm 1$ can be improved approximate 1.86 times, while the SNR was also increased. In further step, we imaged the local magnetic field of nanoparticles. The time-gated detection improved the contrast of magnetic field imaging. Our method can be applied to various types of NV spin manipulation.
Combining with super-resolution microscopy of NV center\cite{chen201501,chen-prapp-2017}, we expect that the high contrast high resolution quantum microscope can be realized for the studies of nanoscience.

\section*{Acknowledgment}
We thank Xi'an SuperMag Bio-Nanotech Co., Ltd for providing us Fe$_{3}$O$_{4}$ nanodisks. This work was supported by National Key Research and
Development Program of China (No. 2017YFA0304504); Anhui Initiative in Quantum Information Technologies (AHY130100); National Natural
Science Foundation of China (Nos. 91536219, 61522508, 11504363, 91850102).

\end{document}